\documentclass[useAMS,usenatbib]{mn2e}

\usepackage{times}
\usepackage{graphicx}
\usepackage{natbib}
\usepackage{multirow}

\usepackage{rotating}  
\usepackage{longtable}
\usepackage{lscape}
\title[The AT20G high angular resolution catalogue]{The AT20G high angular resolution catalogue}

\author[R. Chhetri et
al.] {R. ~Chhetri$^{1,2}$\thanks{email:rchhetri@phys.unsw.edu.au},
R. D. ~Ekers $^{2}$, P. A. ~Jones$^{1}$, R. ~Ricci$^{3}$\\
$^{1}$Department of Astrophysics \& Optics, School of Physics, University of
New South Wales, NSW 2052, Australia\\
$^{2}$Australia Telescope National Facility, CSIRO Astronomy and Space Science,
P.O. Box 76, Epping, NSW
1710, Australia\\
$^{3}$Instituto di Radioastronomia, INAF, Via Gobetti 101,
40129 Bologna,
Italy}

\begin{document}

\maketitle

\begin{abstract}
 We present the high angular resolution catalogue for the Australia Telescope 20 GHz (AT20G) survey, using the high angular resolution 6-km antenna data at the baselines of $\sim$ 4500 m of the Australia Telescope Compact Array (ATCA). We have used the data to produce the visibility catalogue that separates the compact Active Galactic Nuclei (AGNs) from the extended radio sources at the 0.15 arcsec angular scale, corresponding to the linear size scale of 1 kpc at redshifts higher than 0.7. We find the radio population at 20 GHz to be dominated by compact AGNs constituting 77\% of the total sources in the AT20G. We introduce the visibility-spectra diagnostic plot, produced using the AT20G cross-matches with lower frequency radio surveys at 1 GHz (the NRAO-VLA Sky Survey (NVSS) and the Sydney University Molonglo Sky Survey (SUMSS)), that separates the 20 GHz population into distinct sub-populations of the compact AGNs, the compact steep-spectrum sources, the extended AGN-powered sources and extended flat-
spectrum sources. The extended flat-spectrum sources include a local thermal emitting population of high latitude planetary nebulae and also gravitational lens and binary black hole candidates among the AGNs.  We find a smooth transition in properties between the compact CSS sources and the AGN populations.

The visibility catalogue, together with the main AT20G survey, provides an estimate of angular size scales for sources in the AT20G and an estimate of the flux arising from central cores of extended radio sources.
 
The identification of the compact AGNs in the AT20G survey provides high quality calibrators for high frequency radio telescope arrays and VLBI observations.

\end{abstract}


\section{Introduction}

Studies of extragalactic radio sources at radio frequencies below a few gigahertz have led to the differentiation of the observed radio populations of extended sources into two ``Fanaroff and Riley'' classes - the FR-I and the FR-II class sources \citep{Fanaroff1974}. These sources, with steep-spectrum, dominate the observed population up to frequencies of a few gigahertz. At these frequencies \citep[e.g.][]{Condon1998, Mauch2003}, a different population of ``compact'' sources with a flat-spectrum (that may or may not be associated with any steep-spectrum components) start to become significant. These compact sources are now accepted to be the nuclei of active galaxies while the extended sources are giant lobes produced by the outflow of material from the inner compact core, channelled outwards by the jets. The literature is not clear in the nomenclature of these objects, hence, we shall refer to the active compact cores of the galaxies as the active galactic nuclei (AGNs); and when the extended 
components exist we shall refer to such sources as ``AGN powered sources''.

In the models of radio populations the compact sources dominate the population at higher radio frequencies \citep[e.g.][]{de_Zotti2005}. In \citet{Sadler2008}, we have confirmed that the source population at 20 GHz is dominated by AGNs. Identification of compact populations is important to understand the properties of these sources, and to allow refinement in the evolution models for these sources. Evolution studies compared to extended sources will give further insights into the nature of these sources and the Universe. 

To obtain robust results from such studies, it is important to obtain clean samples that represent the population of sources under study. In radio frequencies, empirical indicators such as the two-point spectral index of radio sources have been extensively used to identify compact and extended populations. Since there had been no clear consensus on where the limits of such cut-off should be placed, different studies have used different judgement criteria to select such cut-off. Despite using these indicators, the authors of such studies still need to go through painstaking trouble to ensure a clean sample is obtained. This is especially true of samples of flat-spectrum sources as their variability over time can place these sources on different sides of the cut-off limit when using data 
from different epochs.

Since the flat-spectrum sources are expected to be compact, a selection criterium based on the angular size estimate of sources from a well defined survey can provide a clean separation of these populations. In combination with the spectral index, this forms a very powerful method to classify the different source populations. Radio interferometry techniques in the visibility domain provide robust methods of data analysis since the errors are well understood. We use the visibility analysis of sources in the Australia Telescope 20 GHz (AT20G) survey, using the long baseline data from the Australia Telescope Compact Array (ATCA) to provide an estimate of the angular size. We present the catalogue of source visibility and present some results from this analysis. 

In Section \ref{sec:at20g_survey} we describe the AT20G survey. In Section \ref{sec:visibility_calc} we describe our method of calculation of source visibility. We present the visibility catalogue in Section \ref{sec:catalogue}. We present the discussion and some results in Section \ref{sec:discussion}. We summarise our work in Section \ref{sec:summary}.  

\section{The AT20G Survey}\label{sec:at20g_survey}
The AT20G is a blind survey carried out at a high radio frequency of 20
GHz for all declinations of the southern sky using the Australia Telescope Compact
Array (ATCA). It covers a total area of 20, 086 deg$^2$. It has a galactic
latitude cut-off \mbox{$|b|=1.5$} degrees. The survey has a total of 5890 sources
above the flux density limit of 40 mJy at 20 GHz. The AT20G is the largest blind
survey done at such a high radio frequency. Most sources south of the
declination of -15 degrees have follow-up observations
within a month at 4.8 and 8.6 GHz \citep{Murphy2010}. The sources not observed at 4.8 and 8.6 GHz were a result of
scheduling logistics and do not introduce any spectral bias.

The AT20G survey was done in two steps. The first step was the blind scanning survey between 2004 and 2008, using three antennae of the ATCA in a non-standard compact configuration, with two short baselines of 30.6  metres and one 61.2 metres. The details of the hardware used, the scanning survey and the blind scan catalogue are discussed in \cite{Hancock2011}.

The second part of the survey made the targeted follow-up observations of sources using two 128 MHz wide frequency bands centred at 18.8 GHz and 21.1 GHz, giving the central frequency of 20 GHz. In this paper, we present the analysis of the high angular resolution data from the 20 GHz follow-up observations, discussed in detail below. 

\subsection{The 20 GHz Follow-up}
The 20 GHz targeted follow-up of the sources detected by the blind survey was done using standard ATCA correlator, in hybrid configurations. The details of this follow-up and the survey catalogue are provided in \cite{Murphy2010}, and the properties of the sources in the survey are discussed in \cite{Massardi2011}. The compact hybrid H214 configuration was used for the follow-up (except for the use of H75 configuration to re-observe sources between the declinations of 0 to -15 in August 2008). Since only five antennae of the ATCA are movable along the rails, the hybrid configurations are obtained using antennae 1 to 5 only while antenna 6 (commonly referred to as the `6-km antenna') is fixed at approximately 4500 m from the compact configuration. 

The follow-up observations were made within a short period of time (typically within weeks) after the main blind survey of the sky,  by taking snapshot observations of sources confirmed in the blind survey. The AT20G catalogue released by \cite{Murphy2010} only uses the follow-up data from the compact configurations of the ATCA in order to prevent bias against extended sources introduced by the use of long baselines. The AT20G survey is 100 percent reliable (due to the follow-up observations) and complete to 78 percent above 50 mJy, 86 percent complete above 70 mJy and 93 percent complete above 100 mJy (\citealt{Massardi2011}). In this paper, we present and analyse a catalogue of sources in the AT20G survey using the previously unused 6-km antenna data.

\section{Calculation of Visibility}\label{sec:visibility_calc}

\begin{figure} 
{\includegraphics[width=1\linewidth, height=0.8\linewidth]{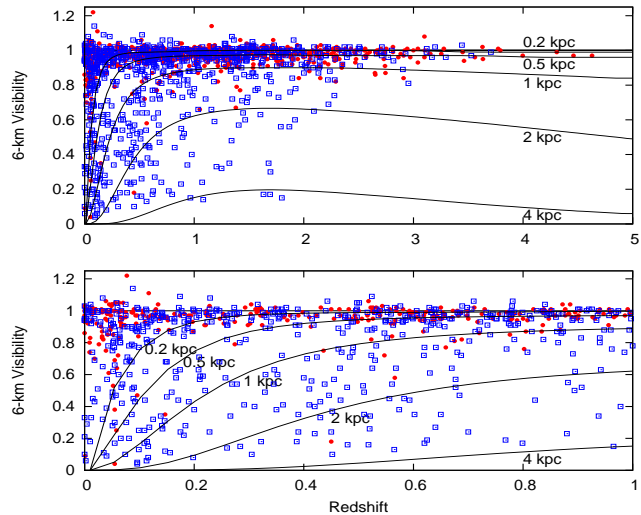}}
        \caption[]{Figure shows the distribution of 6-km visibilities against redshift for AT20G sources. The filled circles are flat-spectrum sources and the open squares are steep-spectrum sources (see Section \ref{sec:spec_index}). Visibility models are drawn at 0.2, 0.5, 1, 2 and 4 kpc. Redshifts are taken from \cite{Mahony2011}}

\label{fig:6kvis-redshift} 
\end{figure}

In the hybrid configurations of the ATCA, the five antennae in the compact configuration and the 6-km antenna provide a maximum baselines between 4300 and 4500 m. These baselines have a factor of 20 greater angular resolution than the compact hybrid configuration used for the main survey. The large gap between the compact configuration of five antennae and the 6-km antenna only provides sparse coverage of interferometer spacings so imaging is not possible. However, non-imaging analysis in the visibility domain can still be used for statistical analysis of source sizes on the sub-arcsec scales. This sub arcsec resolution corresponds to 1 kpc scale for a wide range of redshifts (0.7 - 5) (Figure \ref{fig:6kvis-redshift} and \citet[]{Chhetri2012}). For extragalactic sources this resolution scale is very suitable for the separation of sources into cores and extended jets and lobes. This is also the resolution where the galaxy scale strong gravitational lensing effects are identified (see Section \ref{sec:
discussion}).

At gigahertz radio frequencies, the rapid phase changes due to tropospheric instabilities can cause decorrelation and reduction of visibility amplitude. We can reduce this error by averaging visibility amplitudes instead of the complex visibilities.  This also greatly simplifies our analysis by avoiding the need for phase calibration.

\subsection{Data Processing}
For data processing we used the Multichannel Image Reconstruction, Image Analysis and Display
(Miriad) (\citealt{Sault1995}) software package. We used the AT20G Survey data after the initial data quality
checks implemented in an automated custom analysis pipeline as described by \citet{Murphy2010}.  We then 
calculated the visibilities on the 4500 m baselines in a separate custom pipeline.

\subsubsection{Flagging poor quality data}
The semi-automatic flagging method described in \cite{Murphy2010} removed any data with decorrelations greater than 10 percent and any data with significant attenuation due to atmospheric water vapour opacity. Further, we omitted any hour angle (HA) with $>$ 10\% data flagged as too poor quality for this work. Most of the sources with data flagged using these criteria were re-observed at later epochs.

\subsubsection{Calibration}

Bandpass calibration for data from all epochs of observations was done using the source PKS
1934-638, except for the second 2007 May observations when PKS 0823-500 was used.

The most important error term affecting the data was the multiplicative complex gains terms. We calculated the 6-km visibility by taking the ratio of the correlation amplitudes of long to short baselines measured at the same time and processed in the same way. This cancels some errors and improves the visibility gain calibration.

\subsubsection{Scalar averaging of visibilities}\label{c02:sec:avg_data}

The compact hybrid configuration of the ATCA with five antennae provide 10
short baselines. These short baselines were scalar averaged together to
approximate the maximum amplitude for individual sources (maximum amplitude for any source is obtained at zero spacings of any antenna pair). For unresolved sources, this is a good approximation of the zero spacing visibility. A 3.2 arcsec Gaussian distributed unresolved source will show a 10\% drop in averaged visibility amplitude at the baselines provided by the compact configuration. Sources larger than this are noted as extended in the main AT20G catalogue and are discussed further in Section \ref{sec:src_str}. The visibility amplitudes from the five longer baselines provided by the 6-km antenna were scalar averaged to calculate the long baseline visibility amplitude.

\subsubsection{The 6-km visibility}

We used the data from the 18.8 GHz dataset for this work. Since the visibility of an extended source changes with HA, the individual HA observations were analysed separately. We calculated the ratio of the scalar averaged long baseline visibility amplitudes to the scalar averaged short baseline visibility amplitude to give a normalized visibility value. We obtained the visibility amplitudes using the MIRIAD task UVFLUX. Since this calculation uses the long baselines from the 6-km antenna, we call this ``visibility'' value the ``6-km visibility''. Therefore,

\begin{equation}
	6-km ~visibility= \frac{A_{L}}{A_{S}}
\label{eq:6km-visib}
\end{equation}

where, $A_L$ and $A_S$ stand for scalar averaged visibility amplitudes at long and short baselines respectively.

The normalized quantity shows the ``drop'' in the visibility amplitude for each source in the AT20G catalogue at a baseline of 4500 m. This lowering in visibility amplitude is dependent upon the morphology of the source. The visibility amplitude of sources unaffected by noise will be 1 at all interferometer spacings at which they are unresolved. For resolved sources, the visibility amplitude will be lower at longer baselines, giving a 6-km visibility $<$ 1. Thus, the 6-km visibility measures how well a source is resolved at the baselines of $\sim$4500 metres at 18.8 GHz.

\begin{figure}
       
{\includegraphics[width=1\linewidth]{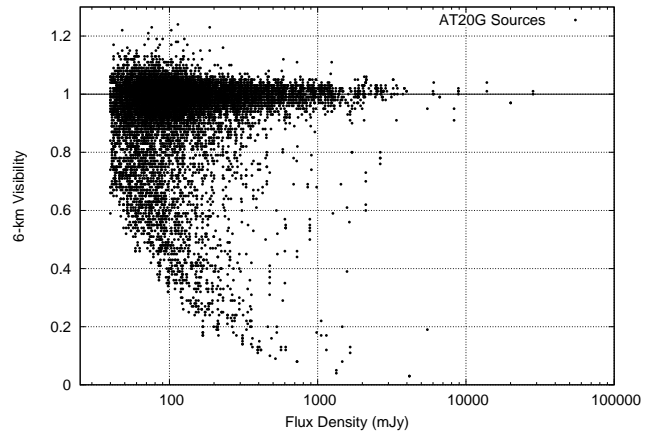}}
        \caption[]{Figure shows the distribution of all 6-km visibilities of AT20G sources against 20 GHz flux density in a log-linear plot. The sharp cut-off on the left is due to the survey flux density limit of 40 mJy. The binning effect seen along the x-axis is due to the precision in the output of Miriad task UVFLUX. The cut-off seen on the lower left is caused by the bias introduced by scalar averaging of visibilities for low signal to noise ratio sources.}

\label{fig:6kvis-all_S20_all_log} 
\end{figure}

We have calculated 6-km visibilities for 5539 (94\%) sources in the AT20G. Each of these sources have between one and a maximum of eight 6-km visibilities at different hour angles. Table \ref{tab:no_6kvis} shows the the distribution of the number of hour angles per source for all sources with 6-km visibilities in the AT20G survey. In Figure \ref{fig:6kvis-all_S20_all_log}, we plot the 6-km visibilities for each source against their 20 GHz flux density from the AT20G survey. 

Our calculation of the visibility amplitudes, using scalar averaging method, is different to the triple product method used to estimate flux densities in the AT20G catalogue. The triple product method provides flux density using the complex triple around all baseline closure triangles and is robust to phase errors \cite[see][for details]{Thompson2001}. While the use of this method is more reliable and less biased for sources which are unresolved on short baselines, it has no simple interpretation for resolved sources. Since our work contains both short and long baselines, it cannot be used to estimate the 6-km visibility.

We compared the use of the 20 GHz flux density (obtained using triple product) against the scalar averaged short baselines amplitude in the denominator of Equations \ref{eq:6km-visib} and \ref{eq:6kvis_bias_corxn}. We found that the calculation of 6-km visibility using the ratio of long to short baselines amplitude, provides better and  individualized corrections for gains giving the best estimate of the visibility amplitude ratio.

\begin{table}
\setlength{\tabcolsep}{3pt}
\begin{small}
\begin{center}
\begin{tabular}{lccccccccc}
\hline
\hline
No. of independent 6-km  & 0 & 1 & 2 & 3 & 4 & 5 & 6 & 7 & 8 \\
visibility observations &  &  &  &  &  &  &  &  &  \\
Number of sources   & 351 & 1316 & 3325 & 330 & 353 & 63 & 140 & 11 & 1\\
\hline

\end{tabular}
\end{center}
\caption[]{Table shows the number of independent HA angles observations and corresponding 6-km visibilities for AT20G sources.}
\label{tab:no_6kvis}
\end{small}
\end{table}

\subsection{Noise Bias} \label{sec:biasCorrxn}

\subsubsection{Noise estimation}
The Gaussian noise in the receiver introduces an additive error to each visibility. We estimated the root mean square (RMS) noise level per 10 second correlator integration for AT20G observations to be 36 mJy (average) using the RMS scatter in the difference in visibility amplitudes (Section \ref{sec:uncertaintyEstimate}). This Gaussian error applies to all visibilities. Since this error is scaled by flux density of the source, it affects the lower flux density sources more than high flux density sources. 

\subsubsection{Bias and its correction}

As the observed signal decreases to the sensitivity limit, the scalar averaged visibility amplitudes approach the RMS noise level for weak sources. This is expected in weak AT20G sources close to the cut-off of 40 mJy. However, since a few hundred individual visibility points (output of each integration) are averaged together, the noise in the resulting visibility is less. Due to the non linear calculation of the visibility amplitude (obtained as the square root of the real and imaginary part of the visibility), the scalar averaged value will be biased by the RMS per sample visibility. This bias is the same for short and long baselines since the noise per sample is the same. Hence, weak sources with zero signals will have 6-km visibility that is artificially closer to 1 for both unresolved and resolved out sources. This effect is seen in lower flux density sources in Figure \ref{fig:6kvis-all_S20_all_log} in the form of a strong cut-off at the bottom left of the plot. The bias thus introduced follows the 
well known 
``Rice 
distribution'' \citep{Taylor1999, Thompson2001} and is similar to that observed in polarization studies of weak sources. In sources with high signal to noise ratio, the errors in the visibility amplitude tends towards a Gaussian distribution in both long and short baselines and the bias is not significant. \\

A \textit{statistical correction} for the bias discussed above can be obtained by removing the averaged noise from the individual visibilities as discussed in \cite[][pp. 319]{Thompson2001}. Then the correction for 6-km visibility is obtained as:

\begin{equation}
	6-km ~visibility^{*}= \frac{\sqrt{(A_{L})^2-\sigma^2}}{\sqrt{(A_{S})^2-\sigma^2}}
\label{eq:6kvis_bias_corxn}
\end{equation}

where,

\begin{tabbing}
	
	$6-km~visibility^{*}$ ~~~~~~~\= is the corrected 6-km visibility,\\
	$A_{L}$ \> is the scalar averaged amplitude of all\\ \> 6-km antenna baselines (in mJy),\\
	$A_{S}$ \> is the scalar averaged amplitude of all \\ \> short baselines (in mJy), and\\
	$\sigma$  \> is the RMS noise per integration \\ \>= 36 mJy.

\end{tabbing}

\begin{figure} 
{\includegraphics[width=1\linewidth]{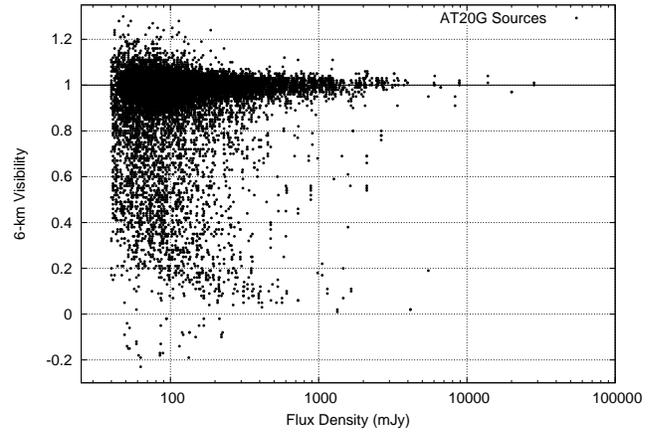}}
        \caption[]{Plot shows 6-km visibility vs. 20 GHz flux density after bias correction shown in Equation \ref{eq:6kvis_bias_corxn}. The correction effectively removes the cut-off in the lower left of Figure \ref{fig:6kvis-all_S20_all_log} due to bias.} 
\label{fig:bias_corrected_6kvis-S20}
\end{figure}

The corrected distribution of 6-km visibility against flux density is plotted in Figure \ref{fig:bias_corrected_6kvis-S20}. Comparing this plot against Figure \ref{fig:6kvis-all_S20_all_log} shows a strong improvement to the distribution of 6-km visibilities in the lower flux density regions, thus showing the successful removal of the bias in weak sources.  All our discussions and analyses in the subsequent sections use the bias corrected 6-km visibility.   

The correction in Equation \ref{eq:6kvis_bias_corxn} can give imaginary values for a very small number of 6-km visibilities (32 out of 11608) for sources with very low to no emission detected in the long baselines. These are shown with negative numbers in the catalogue. These are extended sources with no emission arising from the central ``core'' regions. For such sources, the non-physical negative 6-km visibilities are treated as ``0.00'' in some of the subsequent analyses.

\subsection{Uncertainty estimate}\label{sec:uncertaintyEstimate}

\begin{figure}
{\includegraphics[width=1\linewidth]{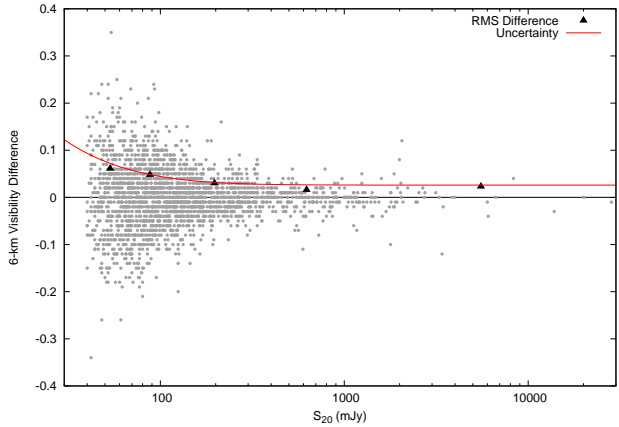}}       
        \caption[]{Plot shows the fit (unbroken line) to the RMS scatter in the difference in the 6-km visibility as a function of 20 GHz flux density to estimate the total uncertainty in the 6-km visibility.} 
\label{fig:uncertaintyEst}
\end{figure}

Sources of visibility uncertainty, both additive and multiplicative, such as antenna pointing errors and receiver gains error combine together to provide uncertainty in the calculated amplitudes and the 6-km visibility. We obtained empirical estimates of uncertainty in the 6-km visibility using the scatter in the 6-km visibility measurements for the multiple hour angle observations. We used only the compact sources identified using the visibility cut-off of 0.86 (see text in Section \ref{sec:sepn-com_ext} and \cite{Chhetri2012}). The total uncertainty in the 6-km visibility can be estimated as $\sigma_{vis}~= \sqrt[]{(\sigma_{n}/S_{20})^2~+~\sigma_{g}}$, where $\sigma_{vis}$ is the total uncertainty in the 6-km visibility, $\sigma_{n}$ is the additive error, $S_{20}$ is the 20 GHz flux density and $\sigma_{g}$ is the multiplicative error. The fit to the RMS scatter in the difference in the visibility of compact sources (Figure \ref{fig:uncertaintyEst}) gives an estimate of 3.6 mJy for additive error and a 2.
5\% gain error.  

Uncertainty in 6-km visibility for each source is presented in column 17 in Table \ref{tab:6km_visib_catalogue}. We note that below 60 mJy, the bias correction in both the numerator and denonimator of Equation \ref{eq:6kvis_bias_corxn} complicates this simple error analysis. The errors will not have a normal distribution and we caution the reader that care will be needed when interpreting these visibilities for the weaker sources.

\section{The 6-km Visibility Catalogue}\label{sec:catalogue}

\subsection{Lower frequency cross-matches}

We made a cross-match between the AT20G sources and sources in the lower frequency surveys covering the entire southern hemisphere sky. The 1.4 GHz NRAO VLA Sky Survey \citep[NVSS,][]{Condon1998} for sources north of $\delta = -40$ degrees, the 0.843 GHz Sydney University Molonglo Sky Survey \citep[SUMSS,][]{Mauch2003} for sources south of $\delta = -30$
degrees and the 0.843 GHz second epoch Molonglo Galactic Plane Survey \citep[MGPS-2,][]{Murphy2007} together cover the total area surveyed by the AT20G. The cross-match of AT20G sources
against the NVSS and the SUMSS (as well as MGPS-2) catalogue provides between one to two spectral points near 1 GHz resulting in flux
densities at two to five frequencies for each source. For convenience, the SUMSS and MGPS-2 catalogues are collectively called the SUMSS catalogue in this paper. The lower frequency points were observed between 2 and 15 years apart compared to the AT20G observations. There is no expected variability in the extended extragalactic sources at these frequencies and less variability in the compact components so we consider these spectral estimates from the low frequency cross-matches reasonably reliable.

\subsection{The 6-km Visibility Catalogue}
We present the total 6-km visibilities in the AT20G as the ``6-km visibility catalogue'' in Tables \ref{tab:6km_visib_catalogue} and \ref{tab:multiple6kvis}. The values are corrected for bias (discussed in Section \ref{sec:biasCorrxn}) according to Equation \ref{eq:6kvis_bias_corxn}. Since the extended sources will be more resolved in the HA with resolution along the major axis the visibilities are sorted in ascending order so that the first listed is closest to the major axis size. We note that this will introduce a small noise bias of the order of the RMS noise error of 3.6 mJy in the visibility amplitude (different to that discussed in Section \ref{sec:biasCorrxn}) when only the minimum value is used. To avoid this bias, we have used all available values for our statistical analyses. The different columns in Table \ref{tab:6km_visib_catalogue} are described below:\\
\textbf{col 1:}   The IAU name for the source.\\
\textbf{col 2:}   The right ascension of the source (J2000)\\
\textbf{col 3:}   The declination of the source (J2000)\\
\textbf{col 4:}   The AT20G 20 GHz flux density (mJy)\\
\textbf{col 5:}   The AT20G 8.6 GHz flux density (mJy)\\
\textbf{col 6:}   The AT20G 4.8 GHz flux density (mJy)\\
\textbf{col 7:}   The NVSS 1.4 GHz flux density (mJy)\\
\textbf{col 8:}   The SUMSS 0.8 GHz flux density (mJy)\\
\textbf{cols 9 - 14:}   Spectral indices calculated between pairs of different frequencies. For sources with flux densities in both NVSS and SUMSS, the NVSS value is used.\\
\textbf{cols 15 - 16:}   The 6-km visibility after bias correction according to Equation \ref{eq:6kvis_bias_corxn}. We have presented 6-km visibilities for each source in the ascending order. \\
\textbf{col 17:}    The uncertainty in all the 6-km visibilities for this source.\\
\textbf{col 18:}    Flags to indicate further information about the source, updated from \cite{Murphy2010}:\\
\textit{n} There are no 6-km visibilites for this source.\\
\textit{v} More than two 6-km visibilities are present. They are listed in Table \ref{tab:multiple6kvis}.\\
\textit{e} The source is flagged as ``extended'' in the AT20G survey and corresponds to angular size greater than $\sim$ 5 arcsec.\\
\textit{h} Source is identified as an HII region.\\
\textit{p} Source is identified as planetary nebula \citep[]{Parker2006, Miszalski2008}.\\
\textit{m} Source is identified as part of Magellanic clouds.\\
\textit{l} Source has no match in the lower frequency surveys (NVSS and SUMSS).\\
\textit{b} Source is flagged as large and extended in the AT20G survey.\\

\pagestyle{empty}

\begin{landscape}
\voffset=1.5in
\hoffset=0.5in
\begin{table*}
\begin{center}
\begin{small}
\begin{tabular}{ccccccccccccccccccccccc}

\hline
IAU Name&RA&DEC&S$_{20}$&S$_{8.6}$&S$_{4.8}$&S$_{1.4}$&S$_{0.8}$&$\alpha_1^{4.8}$&$\alpha_1^{8.6}$&$\alpha_1^{20}$&$\alpha_{4.8}^{8.6}
$&$\alpha_{4.8}^{20}$&$\alpha_{8.6}^{20}$&Vis1&Vis2&Err&Flag\\

&(J2000)&(J2000)&\multicolumn{5}{c}{(mJy)}&&&&&&&&\\ \cline{4-8}

[1]&[2]&[3]&[4]&[5]&[6]&[7]&[8]&[9]&[10]&[11]&[12]&[13]&[14]&[15]&[16]&[17]&[18]\\
\hline
J000012-853919 & 00:00:12.78 & -85:39:19.9 & 98 & 63 & 63 & - & 104 & -0.29 & -0.22 & -0.02 & 0.00 & 0.31 & 0.53 & 0.94 & 0.99 & 0.04 & .\\
J000020-322101 & 00:00:20.38 & -32:21:01.2 & 118 & 315 & 515 & 521 & 322 & -0.01 & -0.28 & -0.56 & -0.84 & -1.03 & -1.17 & 0.97 & 1.06 & 0.04 & .\\
J000105-155107 & 00:01:05.42 & -15:51:07.2 & 297 & 295 & 257 & 348 & - & -0.24 & -0.09 & -0.06 & 0.23 & 0.10 & 0.01 & 0.93 & 0.97 & 0.03 & .\\
J000106-174126 & 00:01:06.31 & -17:41:26.2 & 73 & - & - & 447 & - & - & - & -0.69 & - & - & - & 0.85 & 1.14 & 0.06 & .\\
J000118-074626 & 00:01:18.04 & -07:46:26.8 & 177 & - & - & 208 & - & - & - & -0.06 & - & - & - & 0.91 & 0.99 & 0.03 & .\\
J000124-043759 & 00:01:24.50 & -04:37:59.6 & 50 & - & - & 632 & - & - & - & -0.96 & - & - & - & 0.42 & 0.56 & 0.08 & .\\
J000125-065624 & 00:01:25.59 & -06:56:24.7 & 77 & - & - & 58 & - & - & - & 0.11 & - & - & - & 0.93 & 1.01 & 0.05 & .\\
J000212-215309 & 00:02:12.02 & -21:53:09.9 & 165 & - & - & 370 & - & - & - & -0.31 & - & - & - & 0.72 & 0.80 & 0.03 & .\\
J000221-140643 & 00:02:21.71 & -14:06:43.9 & 48 & - & - & 824 & - & - & - & -1.08 & - & - & - & 0.43 & 0.53 & 0.08 & .\\
J000230-033140 & 00:02:30.60 & -03:31:40.1 & 53 & - & - & 68 & - & - & - & -0.09 & - & - & - & 0.84 & 0.95 & 0.07 & .\\
J000249-211419 & 00:02:49.85 & -21:14:19.2 & 100 & - & - & 122 & - & - & - & -0.08 & - & - & - & 0.87 & 0.92 & 0.04 & .\\
J000252-594814 & 00:02:52.93 & -59:48:14.0 & 71 & 64 & 57 & - & 62 & -0.05 & 0.01 & 0.04 & 0.20 & 0.15 & 0.12 & 0.99 & 1.01 & 0.06 & .\\
J000253-562110 & 00:02:53.65 & -56:21:10.8 & 94 & 229 & 403 & - & 1350 & -0.70 & -0.76 & -0.85 & -0.96 & -1.02 & -1.06 & 0.79 & 1.00 & 0.05 & .\\
J000303-553007 & 00:03:03.45 & -55:30:07.1 & 44 & 48 & 52 & - & 44 & 0.09 & 0.04 & 0.00 & -0.14 & -0.12 & -0.10 & 1.05 & 1.12 & 0.09 & .\\
J000311-544516 & 00:03:11.04 & -54:45:16.8 & 95 & 313 & 552 & - & 1549 & -0.59 & -0.69 & -0.89 & -0.97 & -1.23 & -1.42 & 0.65 & 0.73 & 0.05 & e\\
J000313-590547 & 00:03:13.33 & -59:05:47.7 & 49 & 101 & 151 & - & 506 & -0.70 & -0.69 & -0.74 & -0.68 & -0.79 & -0.86 & 0.84 & 0.88 & 0.08 & .\\
J000316-194150 & 00:03:16.06 & -19:41:50.7 & 76 & 162 & 182 & 232 & - & -0.20 & -0.20 & -0.42 & -0.20 & -0.61 & -0.90 & 0.95 & 1.04 & 0.05 & .\\
J000322-172711 & 00:03:22.05 & -17:27:11.9 & 386 & - & - & 2415 & - & - & - & -0.69 & - & - & - & 0.60 & 0.63 & 0.03 & .\\
J000327-154705 & 00:03:27.35 & -15:47:05.4 & 129 & - & - & 527 & - & - & - & -0.53 & - & - & - & 0.53 & 0.61 & 0.04 & .\\
J000404-114858 & 00:04:04.88 & -11:48:58.0 & 680 & - & - & 459 & - & - & - & 0.15 & - & - & - & 1.01 & -- & 0.03 & .\\
J000407-434510 & 00:04:07.24 & -43:45:10.0 & 199 & 211 & 244 & - & 344 & -0.20 & -0.21 & -0.17 & -0.25 & -0.14 & -0.07 & 0.88 & 0.91 & 0.03 & .\\
J000413-525458 & 00:04:13.97 & -52:54:58.7 & 65 & 98 & 192 & - & 277 & -0.21 & -0.45 & -0.46 & -1.14 & -0.76 & -0.49 & 0.83 & 0.98 & 0.06 & e\\
J000435-473619 & 00:04:35.65 & -47:36:19.0 & 868 & 970 & 900 & - & 933 & -0.02 & 0.02 & -0.02 & 0.13 & -0.03 & -0.13 & 0.98 & 0.99 & 0.03 & .\\
J000505-344549 & 00:05:05.94 & -34:45:49.6 & 131 & 142 & 134 & 92 & 81 & 0.31 & 0.24 & 0.13 & 0.10 & -0.02 & -0.10 & -- & -- & -- & n\\
J000507-013244 & 00:05:07.03 & -01:32:44.6 & 81 & - & - & 71 & - & - & - & 0.05 & - & - & - & 0.96 & 1.03 & 0.05 & .\\
J000518-164804 & 00:05:18.01 & -16:48:04.9 & 142 & - & - & 262 & - & - & - & -0.23 & - & - & - & 0.91 & 0.95 & 0.04 & .\\
J000558-562828 & 00:05:58.32 & -56:28:28.9 & 151 & 376 & 677 & - & 2763 & -0.81 & -0.86 & -0.93 & -1.00 & -1.05 & -1.09 & 0.41 & 0.44 & 0.04 & e\\
J000600-313215 & 00:06:00.47 & -31:32:15.0 & 63 & 53 & 52 & 44 & 48 & 0.13 & 0.10 & 0.14 & 0.03 & 0.13 & 0.21 & 0.91 & -- & 0.06 & .\\
J000601-295549 & 00:06:01.14 & -29:55:49.6 & 97 & 187 & 228 & 88 & - & 0.78 & 0.42 & 0.04 & -0.34 & -0.60 & -0.78 & 0.86 & 0.98 & 0.05 & .\\
J000601-423439 & 00:06:01.95 & -42:34:39.8 & 110 & 259 & 532 & - & 2850 & -0.96 & -1.03 & -1.04 & -1.22 & -1.11 & -1.02 & 0.69 & 0.74 & 0.04 & .\\

\hline \hline
\end{tabular}
\end{small}
\end{center}
\caption[First 30 sources in the AT20G visibility catalogue]{Table shows the first 30 sources in the visibility catalogue of the AT20G survey. Flux densities are in mJy. 6-km visibilities for all sources are listed in columns 15 and 16 in increasing order. 1.4 GHz and 0.8 GHz flux densities are obtained from NVSS and SUMSS catalogues respectively. The full catalogue is presented in the online version. Sources with more than two 6-km visibilities are indicated with ``v'' flag and are listed in Table 3. }
\label{tab:6km_visib_catalogue}
\end{table*}

\end{landscape}

Sources with more than two 6-km visibilities are presented in Table \ref{tab:multiple6kvis}. Column 1 presents the IAU name of the source, columns 2 to 7 present the 6-km visibilities in ascending order and column 8 presents the uncertainty in all the 6-km visibilities for the source.

\begin{table}
\setlength{\tabcolsep}{3pt}
\begin{small}
\begin{center}
\begin{tabular}{cccccccc}
\hline
IAU Name&Vis3&Vis4&Vis5&Vis6&Vis7&Vis8&Err\\
 {[}1{]}&{[}2{]}&{[}3{]}&{[}4{]}&{[}5{]}&{[}6{]}&{[}7{]}&{[}8{]}\\

\hline

J000012-853919 & 1.02 & -- & -- & -- & -- & -- & 0.04\\
J000610-830543 & 0.48 & -- & -- & -- & -- & -- & 0.03\\
J000720-611306 & 1.02 & 1.02 & 1.03 & -- & -- & -- & 0.04\\
J001146-844319 & 0.99 & -- & -- & -- & -- & -- & 0.03\\
J001601-631009 & 0.73 & -- & -- & -- & -- & -- & 0.06\\
J001834-062834 & 0.75 & 0.94 & 0.99 & -- & -- & -- & 0.06\\
J001956-625306 & 0.98 & -- & -- & -- & -- & -- & 0.09\\
J002100-245107 & 0.79 & 0.84 & 0.92 & -- & -- & -- & 0.09\\
J002308-650655 & 1.13 & -- & -- & -- & -- & -- & 0.07\\
J002321-604612 & 1.09 & -- & -- & -- & -- & -- & 0.08\\
J002406-682055 & 0.96 & 1.00 & -- & -- & -- & -- & 0.03\\
J002430-292853 & 0.76 & 0.84 & 0.94 & -- & -- & -- & 0.04\\
J002741-660351 & 1.02 & -- & -- & -- & -- & -- & 0.07\\
J003814-611245 & 1.04 & -- & -- & -- & -- & -- & 0.06\\
J003816-012204 & 1.03 & 1.04 & -- & -- & -- & -- & 0.04\\
J003820-020740 & 0.54 & 0.55 & 0.56 & -- & -- & -- & 0.03\\
J003820-032958 & 1.04 & 1.04 & 1.08 & -- & -- & -- & 0.05\\
J003849-011723 & 1.03 & 1.12 & -- & -- & -- & -- & 0.07\\
J003852-012154 & 1.10 & 1.18 & -- & -- & -- & -- & 0.05\\
J003906-094246 & 1.03 & 1.05 & -- & -- & -- & -- & 0.03\\
J003913-025708 & 1.07 & 1.08 & 1.10 & 1.14 & -- & -- & 0.06\\
J003931-111101 & 0.92 & 0.92 & -- & -- & -- & -- & 0.03\\
J003939-094931 & 1.05 & 1.10 & -- & -- & -- & -- & 0.06\\
J003939-130059 & 1.01 & 1.02 & -- & -- & -- & -- & 0.06\\
J004008-122746 & 1.06 & 1.08 & -- & -- & -- & -- & 0.04\\
J004019-023118 & 0.98 & 1.00 & 1.01 & -- & -- & -- & 0.05\\
J004020-004032 & 0.73 & 0.74 & 0.75 & -- & -- & -- & 0.05\\
J004057-014633 & 1.06 & 1.08 & -- & -- & -- & -- & 0.03\\
J004126-014315 & 0.50 & 0.56 & -- & -- & -- & -- & 0.05\\
J004222-685850 & 1.05 & 1.15 & -- & -- & -- & -- & 0.07\\

\hline

\end{tabular}
\end{center}
\caption[]{Table shows 6-km visibilities for the first 30 sources for which more than two 6-km visibilities are present. Column 8 lists the uncertainty in all 6-km visibilities for each source.}
\label{tab:multiple6kvis}
\end{small}
\end{table}

\section{Discussion}\label{sec:discussion}

\subsection{Separating the compact and extended populations}\label{sec:sepn-com_ext}

\begin{figure} 
{\includegraphics[width=1\linewidth]{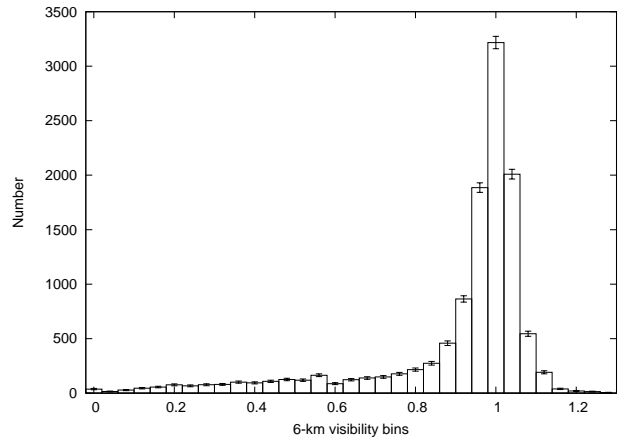}}
        \caption[]{Histogram of 6-km visibility after bias
correction as shown in Equation \ref{eq:6kvis_bias_corxn}. The bin size is 0.04. 
The error bars signify statistical errors in each bin. 6-km visibilities $<$0 are included in the zero bin.} 
\label{histo:bias_corrected_6kvis}
\end{figure}

We have calculated the 6-km visibilities for $>94\%$ of the total sources in the AT20G catalogue. In Figure \ref{histo:bias_corrected_6kvis} we present the histogram of all 6-km visibilities. Unresolved sources have a maximum visibility of 1, so the 6-km visibility values greater than 1 in Figure \ref{histo:bias_corrected_6kvis} are due to noise. This distribution of visibilities $>$ 1 can be used to make a simple empirical estimate of the reliability of our assessment of whether a source is unresolved. While a small contribution can be expected from sources with true 6-km visibilities $<$ 1, the 6-km visibilities with values $>$ 1 are dominated by unresolved sources.

We calculated the 3-sigma value of 0.14 for the distribution of the visibilities for sources with values $>$ 1 \citep[see also][]{Chhetri2012}. Using this, we arrive at the lower limit of 0.86 for the 6-km visibilities of unresolved sources with 3-sigma confidence. Hence, all sources with 6-km visibility $\geq$ 0.86 can be considered compact and sources with 6-km visibility $<$ 0.86 can be considered extended. 77\% of the total sources with 6-km visibility are compact and 23\% are extended, using this definition (Table \ref{tab:no_sources_quadrants}). From the small number of 6-km visibilities $>$ 1.14, only 1.9\% sources would be misclassified. For a source with a Gaussian brightness distribution, the value of 0.86 corresponds to the spatial resolution of 0.15 arcseconds.

The use of a visibility of 0.86 to separate compact and extended sources provides a simple classification of compact and extended sources which has been used in a number of papers analysing the AT20G sources \citep[]{Murphy2010, Massardi2011, Mahony2011, Chhetri2012, Sadler2013}.  Inspection of Figure \ref{fig:bias_corrected_6kvis-S20} makes it clear that this is a conservative estimate for the stronger sources but there will be a small number of misclassified sources at the lowest flux densities.  Analysis of population changes with flux density should include the visibility errors and this will be included in the scope of a subsequent publication.

\subsection{Spectral Index Distribution}\label{sec:spec_index}

We calculated spectral indices\footnote{We use the spectral index definition of $S_{\nu}\propto \nu^{\alpha}$.} between all the frequencies available for our analysis. There is a small number of sources in the overlapping regions of coverage of NVSS and SUMSS that have flux densities measured at both 1.4 and 0.8 GHz respectively. In such cases, the NVSS flux density was used for the calculation of the spectral index. For sources with declinations $\delta >-15$ degrees and other sources without the AT20G 4.8 GHz observations, no lower frequency spectral indices have been calculated. For these sources, the spectral index between 1 and 20 GHz can be used as a proxy spectral index. 

 \begin{figure}
 \begin{center}

{\includegraphics[width=0.9\linewidth]{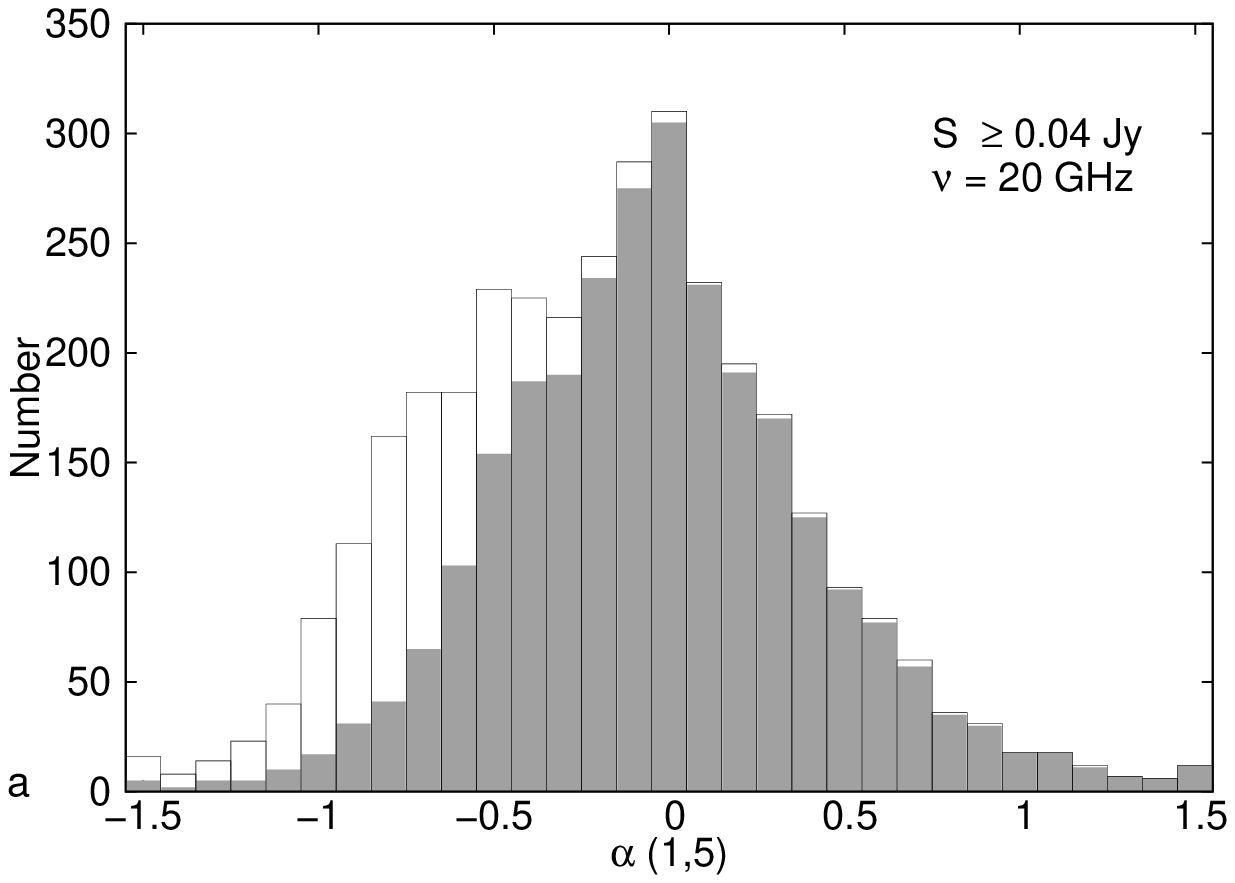}} 
{\includegraphics[width=1\linewidth]{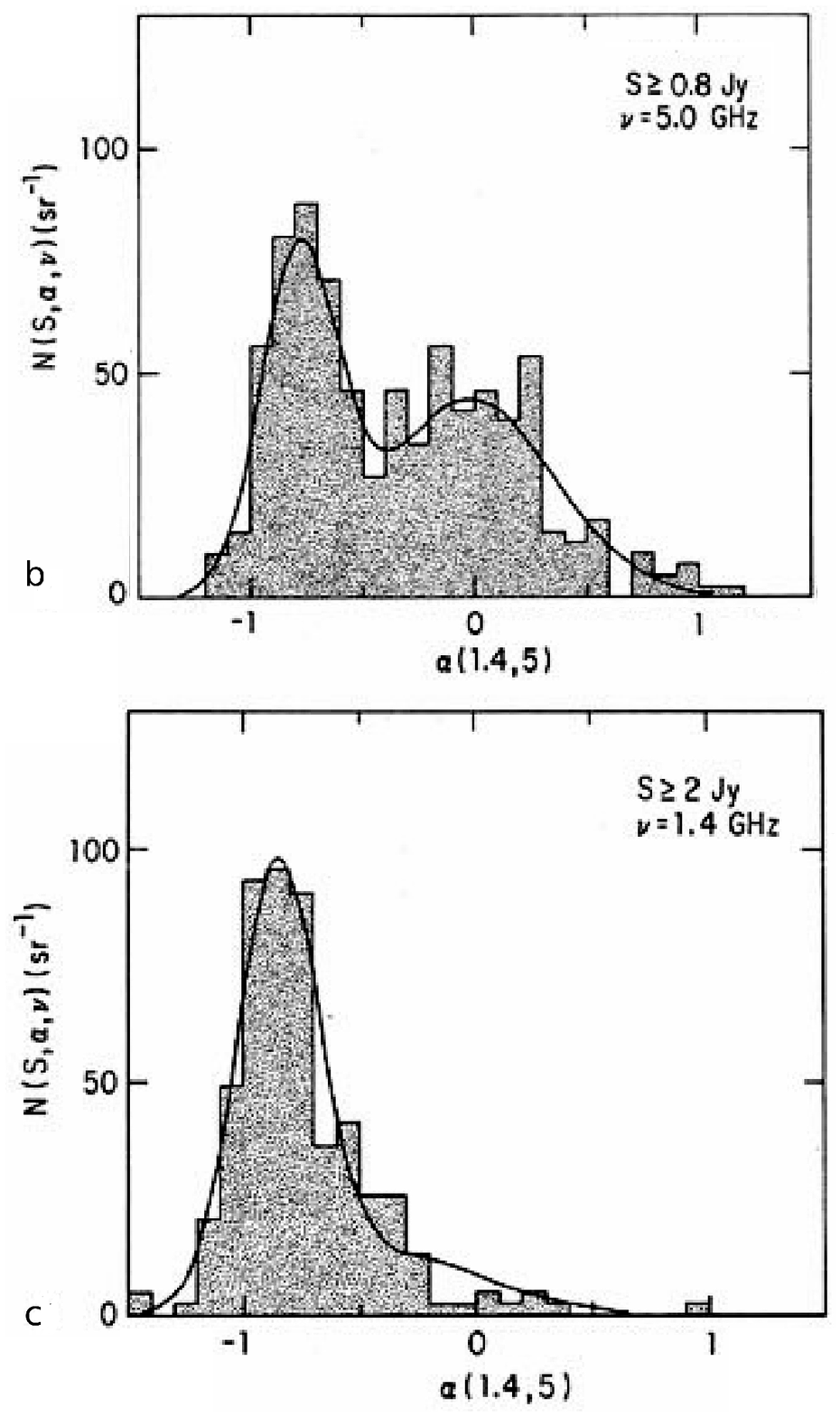}}      
        \caption[]{Plot compares the distribution of spectral index between 1 and 5 GHz for sources selected at different survey frequencies. Plot (a) shows the spectral index distribution of sources in the AT20G. The spectral indices have been calculated between 1 GHz (either NVSS or SUMSS) and AT20G 4.8 GHz frequencies. Shaded part shows the histogram for compact sources. Plots (b) and (c) are produced from \cite[figure 15.7]{Verschuur1988} with changed spectral index definition to $S_{\nu}\propto \nu^{\alpha}$.}  
\label{histo:compareAlpha}
\end{center}

\end{figure}

The distribution of spectral index of the population of sources in a survey depends on the sample selection frequency. At lower frequencies, the population of sources is dominated by steep-spectrum sources. In the gigahertz frequencies, a change from the steep-spectrum population to flat-spectrum is observed. In Figure \ref{histo:compareAlpha} we plot the spectral index distribution of the AT20G sources between 1.4 and 4.8 GHz. A direct comparison of this plot can be made against those observed from 1.4 GHz and 5 GHz in \citet[][chapter 15, figure 15.7]{Verschuur1988}, reproduced as Figure \ref{histo:compareAlpha}(b) and (c). The population at 1.4 GHz is completely dominated by steep-spectrum sources. At 5 GHz there is a bimodal distribution with average spectral index of 0.0 and -0.7 for the flat and the steep-spectrum populations respectively.  This flat spectrum component becomes dominant in our 20 GHz survey and corresponds perfectly to the compact population in Figure \ref{histo:compareAlpha}(a).  This 
is also consistent 
with the expectation for radio source  populations from lower frequency studies (e.g. \cite{de_Zotti2005} and also \cite{Sadler2006}). The compact sources (shaded in Figure \ref{histo:compareAlpha}a) have a surprisingly symmetric distribution of spectral index centred on $\alpha_{1}^{4.8}$ = 0.0. 
\subsection{Source structure} \label{sec:src_str}

With this sparse coverage of interferometer spacings we have only limited information about the source structure but we can determine some key information about the statistics of sources with small components and in some cases we can estimate the size of the small scale components.  The sources with visibility $>$ 0.86 must be smaller than 0.15 arcsec which is less than 1 kpc for all redshift $>$ 0.7 and for closer sources corresponds to scales down to a few hundred parsecs (Figure \ref{fig:6kvis-redshift}b).

The unresolved sources are the cores of active galaxies.  These sources may also have structure on larger scales and in the most extreme cases this could be completely resolved even at the short baselines used in the survey.  Sources with extended structure $>$20\arcsec in size are in this category.  Some of these very large sources are flagged ``b'' in the AT20G catalogue and in Table \ref{tab:6km_visib_catalogue}.  In these cases the observed visibility will be entirely due to the compact component.  For example, Centaurus A is a very extended AGN powered source with large scale structure on a degree scale but the entry only applies to the 6 Jy AGN core.

Sources flagged as extended (e) in the AT20G catalogue and in Table \ref{tab:6km_visib_catalogue} are partially resolved on the short baselines.  These are greater than 5 arcsec in size so if they have significant 6-km visibility they will contain some structure on smaller scales.  In most cases the visibility will be a good estimate of the fraction of the flux density arising from the core, but in some cases it may be unresolved hotspots in the lobes.

The rest of the sources which are resolved on the 6-km antenna baselines but not extended on short baselines will have angular sizes in the range 5\arcsec to 0.15\arcsec.  These could either be multiple lobed sources or core jet sources.  In most cases it would be reasonable to assume that the 6-km visibility is a measure of the fraction of  the flux density arising from the central core.

We present 6-km visibilities from each HA independently. Hence, we expect unresolved sources as well as sources showing isotropic morphological symmetry to show similar values of 6-km visibilities, within errors. For extended sources with asymmetric morphology the lower visibility will be in the position angle closest to the major axis.

A source with multiple components will show beating patterns in the visibility amplitude. The frequency of the beating pattern depends upon the separation between the components while the maxima and the minima in the beats amplitude depends upon the ratio of component amplitudes.  At a given baseline, the 6-km visibility for a resolved double source does not indicate the source size but still provides an upper limit on the component size.

Figure \ref{histo:bias_corrected_6kvis} shows a sharp peak at 1.0 for the visibilities of  the unresolved sources and then a relatively smooth distribution of visibilities for the resolved sources corresponding to angular sizes \mbox{$>$ 0.15\arcsec}.  This is consistent with a smooth distribution in source sizes for the extended source population and a rapid increase in the numbers for the population  of sources \mbox{$<$ 0.15\arcsec}.  We know from VLBI observations that many of these are another order of magnitude smaller than our limit.

\subsection[]{Visibility-spectra source classification}
\begin{figure*}
{\includegraphics[width=1\linewidth]{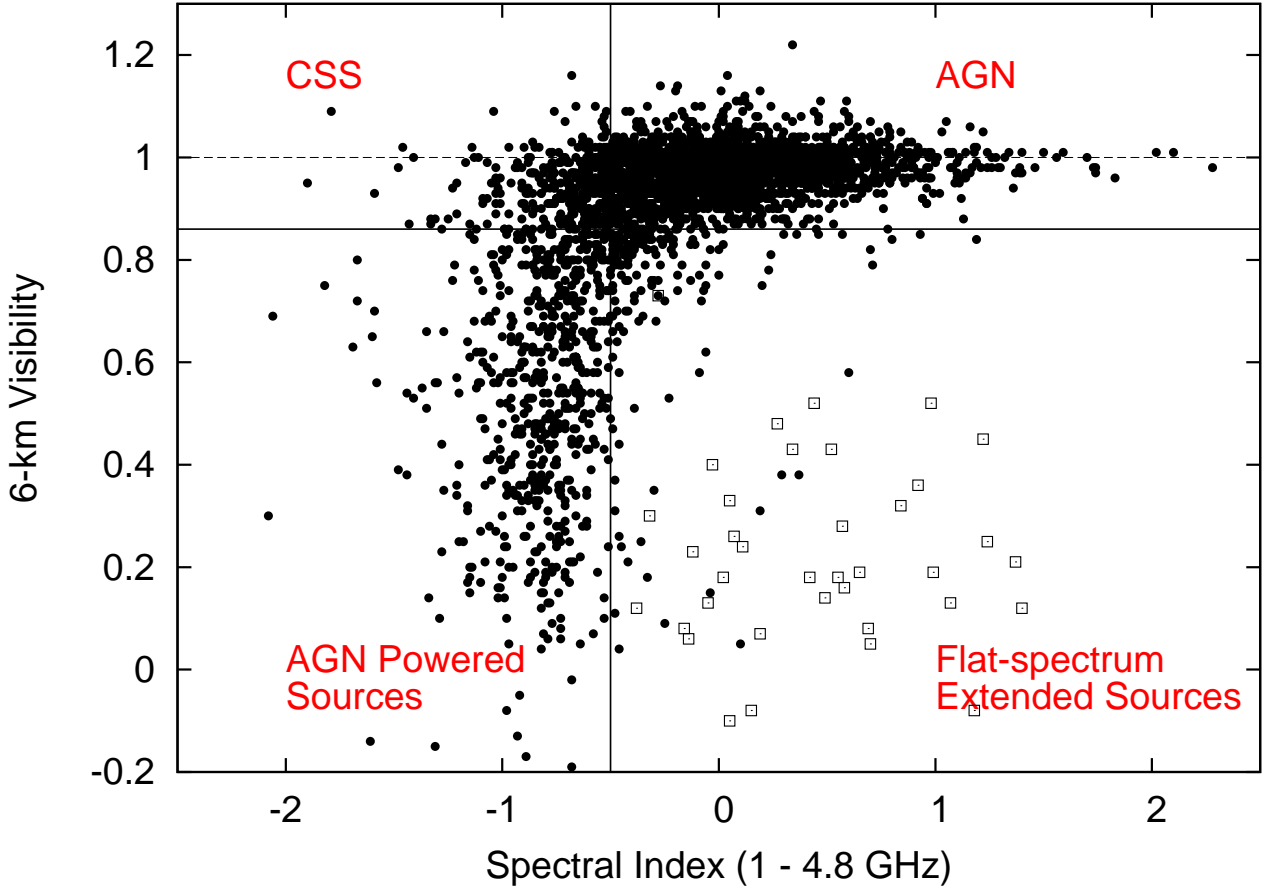}}
        \caption[]{We present the visibility-spectra plot using the lower frequency spectral index for AT20G sources and 6-km visibility. The solid line at 6-km visibility of 0.86 marks the separation between compact and extended populations. A strong correlation between compact and flat-spectrum, and extended and steep-spectrum sources is obvious. A small number of flat-spectrum extended thermal galactic sources detected in the AT20G are marked with open squares (removed for extragalactic analyses). The negative values of 6-km visibilities are a consequence of the bias correction (see text).} 
\label{fig:visib-spectra_main}
\end{figure*}

A direct correlation between the 6-km visibility and spectral index has been discussed in \citet[]{Chhetri2012}. Based on the recommendations therein, a spectral cut-off can be made at $\alpha_1^{4.8}=-0.5$ to separate the total population of sources into flat- and steep-spectrum sources. In Figure \ref{fig:visib-spectra_main}, we plot the lowest 6-km visibility versus spectral index for a sub-population of 3464 AT20G sources, this includes a total of 37 thermal sources. These thermal sources are removed for all extragalactic source analysis. This sub-population  of AT20G sources have both the lower frequency spectral indices ($\alpha_1^{4.8}$) and 6-km visibilities. The two population Kolmogorov-Smirnov (K-S) test for 6-km visibilities for this subgroup and the total 6-km visibilities shows a chance probability of null hypothesis of being ``drawn from the same population'' of $\sim$67\%. Hence, this sub-population is a good representative of the parent population.We use the lower frequency (1 - 4.8 GHz) 
spectral index for this 
analysis rather than a higher frequency spectral index because it separates the extended and 
compact emission populations more effectively.  As discussed in \cite{Chhetri2012} at higher frequencies all sources tend to have similar steep spectra. 

A strong correlation between compact and flat-spectrum, and extended and steep-spectrum sources is obvious in the visibility spectra plot (Figure \ref{fig:visib-spectra_main}). Based on the spectral index transition at -0.5 and the compact source transition at 6-km visibility = 0.86, it is possible to divide Figure \ref{fig:visib-spectra_main} into four quadrants that are occupied by sources with very different properties. The number of sources in each quadrant of the visibility spectra are given in Table \ref{tab:no_sources_quadrants}. We explore these four quadrants below. \\

\begin{table}

\begin{center}
  \begin{tabular}{l|rrr}
    \hline
    \textbf{Spectrum}& \textbf{Compact \textit{(\%)}} & \textbf{Extended \textit{(\%)}} & \textbf{Total \textit{(\%)}}\\
    \hline
    \textbf{All} & 4259 \textit{(76.9)} & 1280 \textit{(23.1)} & \textbf{5539}\\
    \hline
    \textbf{Flat-} & 2335 \textit{(68.1)} & 151 \textit{(4.4)} & 2486 \textit{(72.5)}\\
    \textbf{Steep-} &  374 \textit{(10.9)} & 567 \textit{(16.5)} & 941 \textit{(27.5)}\\
    \textbf{Total} & 2709 \textit{(79.0)} & 718 \textit{(20.9)} & \textbf{3427 }\\
    \hline \hline
  \end{tabular}
  \end{center}
  \caption{Table shows the number of compact and extended sources in the AT20G survey using the 6-km visibility classification. The first row shows numbers for all AT20G sources with 6-km visibilities. The bottom part shows the number of extragalactic sources in each quadrant of Figure \ref{fig:visib-spectra_main} only. The percentages are calculated for the 3427 extragalactic sources that have both lower frequency spectral indices and 6-km visibilities (see text).}
  \label{tab:no_sources_quadrants}
\end{table}

\subsection*{Compact and flat-spectrum sources:}
The top right quadrant of the plot is occupied by the compact and flat-spectrum sources. These sources form only a small fraction of total population in the lower frequency surveys such as the NVSS or the SUMSS but constitute the most significant fraction (68\%) of the total population at 20 GHz. The visibility cut-off of 0.86 corresponds to an angular size of 0.15 arcsec and for a wide range of redshifts (Figure \ref{fig:6kvis-redshift}) this corresponds to a linear size less than 1kpc.  This includes both the blazar population with beamed and flaring synchrotron emission and the young Gigahertz Peaked Spectrum (GPS) and their High Frequency Peaking (HFP) extension \citep{O_Dea1998}. Most AT20G sources in this quadrant have spectra with a broad peak between 1 and 20 GHz \citep[]{Massardi2011b, Chhetri2012, Bonaldi2013} but we can not tell whether they are GPS/HFP sources or blazars.  However, we note that the \cite{Sadler2006} and \cite{Massardi2011b} variability study of a subsamples of the AT20G sources 
shows 
only 40\% have variability of more than 10\% in a nine months to one year time period. Hence, there are many compact flat-spectrum sources in this quadrant which are not strongly variable in the one year time scale. However, these sources may be affected by variability in longer time scales (e.g. \cite{Hovatta2007} suggest a larger variability in a $\sim$ 6 year time period for different classes of AGNs). This population of compact sources provides direct applications as phase reference calibrators for existing as well as future instruments such as the ATCA, the Atacama Large Millimeter Array (ALMA) and very large baseline interferometry (VLBI).

\subsection*{Compact and steep-spectrum sources:}
The top left quadrant is occupied by sources that are compact and show steep-spectum at lower frequencies. Only about 11\% of the total population is found in this quadrant. We find three classes of source in this quadrant.  A few are simply compact flat spectrum sources that have diffuse steep spectrum emission on scales greater than 60 arcseconds, included in the NVSS or SUMSS flux density but are completely resolved out in the 20 GHz ATCA observations.  As described in section \ref{c02:sec:avg_data} this visibility catalogue refers to the size of the compact component in the nucleus not the total source.  \mbox{Figure \ref{fig:visib-spectra_CSS}} shows this quadrant with the sources with steep spectra at all frequencies differentiated. These are the Compact Steep-spectrum Sources (CSS) discussed by \cite{O_Dea1998} and thought to be an early stage of AGN evolution. These sources with steep-spectra at all frequencies constitute 5\% (143 sources) of the total compact population and show median spectral 
indices of $\alpha_{1}^{4.8}=-0.
66, \alpha_
{4.8}^{8.6}=-0.81$ and $\alpha_{8.6}^{20}=-0.97$. A very small number of these sources might be very high redshift compact sources in which the general spectral steepening has been redshifted down to GHz frequencies \citep[]{Chhetri2012}. \\\\
{\textbf{The CSS-AGN transition region:}\\
In Figure \ref{fig:visib-spectra_CSS}, we select sources that have steep-spectral indices at higher frequencies ($\alpha_{4.8}^{8.6}\leq-0.5$ and $\alpha_{8.6}^{20}\leq-0.5$). The CSS sources identified above (left of vertical line at $\alpha_{4.8}^{8.6}=-0.5$) merge smoothly into the class of flat-spectrum AGNs. Compact sources at high redshifts showing spectral steepening and GPS sources at later stages of evolution are expected to populate the region between the two vertical lines in Figure \ref{fig:visib-spectra_CSS}. Sources to the right of the vertical line at $\alpha_{4.8}^{8.6}=-0.5$ have a gigahertz peak in their spectrum but appear less connected to the bulk of the CSS and AGN population.

\subsection*{Extended and steep-spectrum sources:}
The lower left quadrant of the plot is occupied by extended sources showing steep-spectrum. These are mostly AGN powered extended
radio sources with the radio emission arising from steep-spectrum jet(s) and/or lobe(s). These sources may still contain a weak core
component. Sources such as the canonical extended radio source Cygnus A are expected to occupy this quadrant. These sources that dominate the lower frequency surveys such as the NVSS or the SUMSS constitute only about 17\% of the total sources at 20 GHz. These sources show steep-spectrum at higher frequencies as well. But for a fraction of these sources, the spectrum becomes flatter when the compact component dominates the total spectra \citep{Chhetri2012}.

\subsection*{Extended and flat-spectrum sources:}
The lower right quadrant of the plot is occupied by sources that are flat-spectrum but extended. Thermal sources such as galactic HII
regions including planetary nebulae are also expected to occupy this quadrant. We identify 33 planetary nebulae \citep[]{Parker2006, Miszalski2008} and four HII regions. These  galactic thermal sources (including three extragalactic HII region in the Large Magellanic Cloud; to be presented in future publication. For any more distant galaxies, the thermal free-free emission can be expected to fall below the detection threshold of the AT20G survey.) are identified in the plot and have been removed for analyses of extragalactic populations. Flat-spectrum compact sources (AGNs) in the top right quadrant, described above, can be brought down to this quadrant by the effect of strong gravitational lensing. Any binary AGN systems where each individual AGN component has a flat-spectrum can also be found in this quadrant (Chhetri et al. in prep). Our follow-up to find the gravitational lenses using full imaging at 7mm (Chhetri et al. in prep) have shown that some of these are core-jet sources with flat-spectrum cores 
strong enough to dominate the spectral index but with sufficient extended structure to significantly reduce the 6-km visibility.  They are 
mainly found in the top left corner of the quadrant. We also identify the low redshift galaxy NGC 253 with $\alpha_{1}^{4.8}=-0.25$ and 6-km visibility of 0.09 in this quadrant, the only source where the 20 GHz emission appears to arise from a central starburst \citep{Sadler2013}. We remind the readers that sources with low 6-km visibilities are resolved and their 20 GHz flux densities may be under represented in the AT20G catalogue.

 \begin{figure}
{\includegraphics[width=1\linewidth]{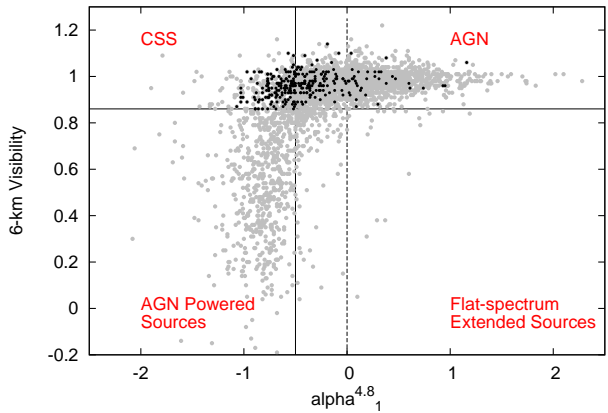}}       
        \caption[]{We identify compact sources that have steep spectrum at frequencies above 4.8 GHz. The sources, represented by dark filled circles, on the right hand side of the vertical dotted line at $\alpha_1^{4.8} = 0.0$ have their spectra peaking at gigahertz frequencies. Dark filled circles with $\alpha_1^{4.8} \leq -0.5$ are CSS sources. } 
\label{fig:visib-spectra_CSS}
\end{figure}

\section{Summary} \label{sec:summary}

We present a catalogue of interferometric visibilities on the sub arcsec scale for the AT20G survey. The normalized visibility amplitudes, termed 6-km visibility, have been calculated as the ratio of visibility amplitudes of the long and the short baselines of the ATCA. The catalogue contains visibilities for $>$ 94\% of the total number of sources in the AT20G. Sources have between 1 and 8 independent 6-km visibilities. 

The catalogue provides a clean separation between the compact and extended sources at the 6-km visibility of 0.86 corresponding to the angular scale of 0.15 arcseconds. 77\% of the total population are found to be compact AGNs and 23\% are extended AGN powered sources at 20 GHz. The 6-km visibility of 0.86 corresponds to 1 kpc in linear size scale at redshift $>$0.7 and separates the AGN population.

The spectral indices obtained from the cross-matches with lower frequency NVSS and SUMSS catalogues show a strong correlation between compact and flat-spectrum, and extended and steep-spectrum sources. The compact sources show a very symmetric distribution of spectral index centered on $\alpha_1^{4.8} = 0.0$.  We present the visibility-spectra plot that separates the total population into four different populations of AGNs (including the blazars and GPS sources), compact and steep-spectrum sources (including the CSS sources), AGN powered sources and an interesting class of extended flat-spectrum sources. We note a smooth transition in spectral index between the CSS sources and the AGNs in the visibility-spectra plot, while that with the GPS sources is noted to be less clear. We identify a total of 33 planetary nebulae and four HII regions in the extended flat-spectrum population, including three extragalactic HII regions in the Large Magellanic Cloud. Further, we identify this as the 
population where instances of galaxy scale gravitational lensing and supermassive binary black hole systems may be found.

The use of the 6-km visibility catalogue, along with the AT20G catalogue, makes it possible to obtain limits on the angular size scale for sources in the AT20G. For sources in the 0.5\arcsec - 5\arcsec size scales, the 6-km visibility can be used to make good estimate of the flux from the central cores.

The visibility catalogue provides an excellent method of selecting different populations of extragalactic sources at subarcsec angular resolution scale in the relatively unexplored high radio frequencies. This provides particularly important constraints on extragalactic source population models which are  being used to predict the properties of the radio sky at the much lower (Square Kilometer Array (SKA) level) flux densities since it is the high frequency sources that will be redshifted into the lower frequency surveys \citep{Mahony2011}. 

In addition to its value separating the high frequency source population, there is the direct applications as high quality compact phase reference calibrators for existing and upcoming instruments such as the ATCA, ALMA and VLBI.

\section*{Acknowledgements}

R.C. would like to thank Tara Murphy for providing cross-matching scripts and S. Chhetri for other support. We thank M. Cohen and B. Miszalski for their expertise and assistance in identifying the galactic thermal sources.

\bibliographystyle{scemnras}
\bibliography{references}

\begin{thebibliography}{23}
\expandafter\ifx\csname natexlab\endcsname\relax\def\natexlab#1{#1}\fi

\bibitem[{{Bonaldi} {et~al.}(2013){Bonaldi}, {Bonavera}, {Massardi}, \& {De
  Zotti}}]{Bonaldi2013}
{Bonaldi}, A., {Bonavera}, L., {Massardi}, M., {De Zotti}, G. 2013, \mnras,
  428, 1845

\bibitem[{{Chhetri} {et~al.}(2012){Chhetri}, {Ekers}, {Mahony}, {Jones},
  {Massardi}, {Ricci}, \& {Sadler}}]{Chhetri2012}
{Chhetri}, R., {Ekers}, R.~D., {Mahony}, E.~K., {Jones}, P.~A., {Massardi}, M.,
  {Ricci}, R., {Sadler}, E.~M. 2012, \mnras, 422, 2274

\bibitem[{{Condon} {et~al.}(1998){Condon}, {Cotton}, {Greisen}, {Yin},
  {Perley}, {Taylor}, \& {Broderick}}]{Condon1998}
{Condon}, J.~J., {Cotton}, W.~D., {Greisen}, E.~W., {Yin}, Q.~F., {Perley},
  R.~A., {Taylor}, G.~B., {Broderick}, J.~J. 1998, \aj, 115, 1693

\bibitem[{{de Zotti} {et~al.}(2005){de Zotti}, {Ricci}, {Mesa}, {Silva},
  {Mazzotta}, {Toffolatti}, \& {Gonz{\'a}lez-Nuevo}}]{de_Zotti2005}
{de Zotti}, G., {Ricci}, R., {Mesa}, D., {Silva}, L., {Mazzotta}, P.,
  {Toffolatti}, L., {Gonz{\'a}lez-Nuevo}, J. 2005, \aap, 431, 893

\bibitem[{{Fanaroff} \& {Riley}(1974)}]{Fanaroff1974}
{Fanaroff}, B.~L. {Riley}, J.~M. 1974, \mnras, 167, 31P

\bibitem[{{Hancock} {et~al.}(2011){Hancock}, {Roberts}, {Kesteven}, {Ekers},
  {Sadler}, {Murphy}, {Massardi}, {Ricci}, {Calabretta}, {de Zotti}, {Edwards},
  {Ekers}, {Jackson}, {Leach}, {Phillips}, {Sault}, {Staveley-Smith},
  {Subrahmanyan}, {Walker}, \& {Wilson}}]{Hancock2011}
{Hancock}, P.~J., {et~al.} 2011, Experimental Astronomy, 32, 147

\bibitem[{{Hovatta} {et~al.}(2007){Hovatta}, {Tornikoski}, {Lainela}, {Lehto},
  {Valtaoja}, {Torniainen}, {Aller}, \& {Aller}}]{Hovatta2007}
{Hovatta}, T., {Tornikoski}, M., {Lainela}, M., {Lehto}, H.~J., {Valtaoja}, E.,
  {Torniainen}, I., {Aller}, M.~F., {Aller}, H.~D. 2007, \aap, 469, 899

\bibitem[{{Mahony} {et~al.}(2011){Mahony}, {Sadler}, {Croom}, {Ekers},
  {Bannister}, {Chhetri}, {Hancock}, {Johnston}, {Massardi}, \&
  {Murphy}}]{Mahony2011}
{Mahony}, E.~K., {et~al.} 2011, \mnras, 417, 2651

\bibitem[{{Massardi} {et~al.}(2011{\natexlab{a}}){Massardi}, {Bonaldi},
  {Bonavera}, {L{\'o}pez-Caniego}, {de Zotti}, \& {Ekers}}]{Massardi2011b}
{Massardi}, M., {Bonaldi}, A., {Bonavera}, L., {L{\'o}pez-Caniego}, M., {de
  Zotti}, G., {Ekers}, R.~D. 2011{\natexlab{a}}, \mnras, 415, 1597

\bibitem[{{Massardi} {et~al.}(2011{\natexlab{b}}){Massardi}, {Ekers}, {Murphy},
  {Mahony}, {Hancock}, {Chhetri}, {de Zotti}, {Sadler}, {Burke-Spolaor},
  {Calabretta}, {Edwards}, {Ekers}, {Jackson}, {Kesteven}, {Newton-McGee},
  {Phillips}, {Ricci}, {Roberts}, {Sault}, {Staveley-Smith}, {Subrahmanyan},
  {Walker}, \& {Wilson}}]{Massardi2011}
{Massardi}, M., {et~al.} 2011{\natexlab{b}}, \mnras, 412, 318

\bibitem[{{Mauch} {et~al.}(2003){Mauch}, {Murphy}, {Buttery}, {Curran},
  {Hunstead}, {Piestrzynski}, {Robertson}, \& {Sadler}}]{Mauch2003}
{Mauch}, T., {Murphy}, T., {Buttery}, H.~J., {Curran}, J., {Hunstead}, R.~W.,
  {Piestrzynski}, B., {Robertson}, J.~G., {Sadler}, E.~M. 2003, \mnras, 342,
  1117

\bibitem[{{Miszalski} {et~al.}(2008){Miszalski}, {Parker}, {Acker}, {Birkby},
  {Frew}, \& {Kovacevic}}]{Miszalski2008}
{Miszalski}, B., {Parker}, Q.~A., {Acker}, A., {Birkby}, J.~L., {Frew}, D.~J.,
  {Kovacevic}, A. 2008, \mnras, 384, 525

\bibitem[{{Murphy} {et~al.}(2007){Murphy}, {Mauch}, {Green}, {Hunstead},
  {Piestrzynska}, {Kels}, \& {Sztajer}}]{Murphy2007}
{Murphy}, T., {Mauch}, T., {Green}, A., {Hunstead}, R.~W., {Piestrzynska}, B.,
  {Kels}, A.~P., {Sztajer}, P. 2007, \mnras, 382, 382

\bibitem[{{Murphy} {et~al.}(2010){Murphy}, {Sadler}, {Ekers}, {Massardi},
  {Hancock}, {Mahony}, {Ricci}, {Burke-Spolaor}, {Calabretta}, {Chhetri}, {de
  Zotti}, {Edwards}, {Ekers}, {Jackson}, {Kesteven}, {Lindley}, {Newton-McGee},
  {Phillips}, {Roberts}, {Sault}, {Staveley-Smith}, {Subrahmanyan}, {Walker},
  \& {Wilson}}]{Murphy2010}
{Murphy}, T., {et~al.} 2010, \mnras, 402, 2403

\bibitem[{{O'Dea}(1998)}]{O_Dea1998}
{O'Dea}, C.~P. 1998, \pasp, 110, 493

\bibitem[{{Parker} {et~al.}(2006){Parker}, {Acker}, {Frew}, {Hartley},
  {Peyaud}, {Ochsenbein}, {Phillipps}, {Russeil}, {Beaulieu}, {Cohen},
  {K{\"o}ppen}, {Miszalski}, {Morgan}, {Morris}, {Pierce}, \&
  {Vaughan}}]{Parker2006}
{Parker}, Q.~A., {et~al.} 2006, \mnras, 373, 79

\bibitem[{{Sadler} {et~al.}(2013){Sadler}, {Ekers}, {Mahony}, {Mauch}, \&
  {Murphy}}]{Sadler2013}
{Sadler}, E.~M., {Ekers}, R.~D., {Mahony}, E., {Mauch}, T., {Murphy}, T. 2013,
  arXiv:astro-ph/1304.0268

\bibitem[{{Sadler} {et~al.}(2008){Sadler}, {Ricci}, {Ekers}, {Sault},
  {Jackson}, \& {de Zotti}}]{Sadler2008}
{Sadler}, E.~M., {Ricci}, R., {Ekers}, R.~D., {Sault}, R.~J., {Jackson}, C.~A.,
  {de Zotti}, G. 2008, \mnras, 385, 1656

\bibitem[{{Sadler} {et~al.}(2006){Sadler}, {Ricci}, {Ekers}, {Ekers},
  {Hancock}, {Jackson}, {Kesteven}, {Murphy}, {Phillips}, {Reinfrank},
  {Staveley-Smith}, {Subrahmanyan}, {Walker}, {Wilson}, \& {de
  Zotti}}]{Sadler2006}
{Sadler}, E.~M., {et~al.} 2006, \mnras, 371, 898

\bibitem[{{Sault} {et~al.}(1995){Sault}, {Teuben}, \& {Wright}}]{Sault1995}
{Sault}, R.~J., {Teuben}, P.~J., {Wright}, M.~C.~H. 1995, in Astronomical
  Society of the Pacific Conference Series, Vol.~77, Astronomical Data Analysis
  Software and Systems IV, ed. R.~A. {Shaw}, H.~E. {Payne}, J.~J.~E. {Hayes},
  433

\bibitem[{{Taylor} {et~al.}(1999){Taylor}, {Carilli}, {Perley}, \&
  {G.~B.~Taylor, C.~L.~Carilli, \& R.~A.~Perley}}]{Taylor1999}
{Taylor}, G.~B., {Carilli}, C.~L., {Perley}, R.~A., {G.~B.~Taylor,
  C.~L.~Carilli, \& R.~A.~Perley}, eds. 1999, Astronomical Society of the
  Pacific Conference Series, Vol. 180, {Synthesis Imaging in Radio Astronomy
  II}

\bibitem[{{Thompson} {et~al.}(2001){Thompson}, {Moran}, \&
  {Swenson}}]{Thompson2001}
{Thompson}, A.~R., {Moran}, J.~M., {Swenson}, Jr., G.~W. 2001, {Interferometry
  and Synthesis in Radio Astronomy, 2nd Edition}, ed. {Thompson, A.~R., Moran,
  J.~M., \& Swenson, G.~W., Jr.} (John Wiley \& sons, Inc)

\bibitem[{{Verschuur} \& {Kellermann}(1988)}]{Verschuur1988}
{Verschuur}, G.~L. {Kellermann}, K.~I. 1988, {Galactic and extra-galactic radio
  astronomy} (Berlin: Springer)

\end{thebibliography}

\bigskip
This document has been typeset from a \TeX/\LaTeX  file prepared by the
author.
\end{document}